\documentclass[journal]{IEEEtran}

\usepackage{xcolor,soul,framed} 

\colorlet{shadecolor}{yellow}
\usepackage[pdftex]{graphicx}
\graphicspath{{../pdf/}{../jpeg/}}
\DeclareGraphicsExtensions{.pdf,.jpeg,.png}

\usepackage[cmex10]{amsmath}
\usepackage{array}
\usepackage{mdwmath}
\usepackage{mdwtab}
\usepackage{eqparbox}
\usepackage{url}
\usepackage{amssymb}
\usepackage{booktabs}
\usepackage{caption}
\usepackage{subcaption}
\usepackage[numbers,square]{natbib}
\usepackage{amsmath}
\usepackage{enumerate}
\newtheorem{example}{Example}
\newtheorem{theorem}{Theorem}
\newtheorem{corollary}{Corollary}

\newtheorem{lem}{Lemma}
\newtheorem{definition}{Definition}
\newtheorem{assum}{Assumption}
\begin{document}
\bstctlcite{IEEEexample:BSTcontrol}
    \title{Online coalitional games for real-time payoff distribution with applications to energy markets}
  \author{Aitazaz Ali Raja and
     Sergio Grammatico
\thanks{Aitazaz Ali Raja and Sergio Grammatico are with Delft Center for Systems and Control, TU Delft, The Netherlands. (e-mail addresses: a.a.raja@tudelft.nl; s.grammatico@tudelft.nl).}
  \thanks{This work was partially supported by NWO under research project P2P-TALES (grant n. 647.003.003) and the ERC under research project COSMOS, (802348).}
}  
\maketitle
\begin{abstract}


Motivated by the markets operating on fast time scales, we present a framework for online coalitional games with time-varying coalitional values and propose real-time payoff distribution mechanisms. Specifically, we design two online distributed algorithms to track the Shapley value and the core, the two most widely studied payoff distribution criteria in coalitional game theory. We show that the payoff distribution trajectory resulting from our proposed algorithms converges to a neighborhood of the time-varying solutions. We adopt an operator-theoretic perspective to show the convergence of our algorithms. Numerical simulations of a real-time local electricity market and cooperative energy forecasting market illustrate the performance of our algorithms:  {the difference between online payoffs and static payoffs (Shapley and the core) to the participants is little; online algorithms considerably improve the scalability of the mechanism with respect to the number of market participants.} 

\end{abstract}


%
\IEEEpeerreviewmaketitle

\section{Introduction} \label{sec: intro}

 A technological transformation is currently underway converting key infrastructures, such as power grids, commerce, and trading platforms, into highly dynamic complex systems. In these domains, predictive decision-making and operational planning under uncertainty traditionally rely on forecasts. The reliability of a forecast generally decreases as the lead time increases, especially for systems operating in  highly dynamic environments. Thus, acting closer to the time of occurrence of an event decreases the chance of inaccurate or erroneous decision-making. Alongside, the sweeping technological advances across sectors like communication, sensing, data acquisition, and computation are making time-ahead decision-making a dormant approach. Therefore, we need methodologies and mechanisms that make use of real-time data streams and respond to the fast dynamics of the underlying system via online decision-making \cite{dall2020optimization}.

Among the systems operating in highly dynamic environments, here we focus on real-time markets. The adoption of real-time markets has shown significant potential in the power system sector \cite{wang2015review}. In particular, the increased presence of distributed energy resources (DERs) and demand response (DR) programs on the consumer side allow system operators to utilize them for providing demand-supply balancing services in real time \cite{pineda2013using}. Unlike conventional generators, the response time of DERs and DR fulfills the operational requirements of participation in real-time balancing markets.  In   \cite{guo2021online}, the authors build a model for the real-time operation of a recent market paradigm, i.e., peer-to-peer (P2P) markets. P2P markets envision a bilateral trade of renewable energy among small prosumers. As the accuracy of forecasts can be improved by decreasing the lead time \cite{nielsen2006short}, market clearing closer to the time of delivery can mitigate the possible imbalance caused by the uncertainty associated with RES. In both balancing and P2P markets, the key enabling feature is the computational speed of the clearing mechanism. The mechanisms for such markets require a large amount of information exchange and execution of negotiation processes. Consequently, in the context of real-time markets, the computational time for market clearing can be higher than the  gap between two market instances. 

Another marketplace operating in a dynamic environment that has recently gained a lot of interest from both academia and industry, is the data market \cite{bergemann2019markets}. With the emergence of machine learning across all business and social sectors, the need for quality training data has grown enormously. One way to ensure data availability is by creating a market that compensates data providers. Various structures and mechanisms are proposed in the literature for data markets including bilateral exchange of data \cite{rasouli2021data} and a regression market framework for wind power forecasting \cite{han2021trading}. In general, assigning a value to a particular data set among many is inherently a combinatorial problem. The authors in \cite{ghorbani2019data} address the problem of data valuation for a specified machine learning algorithm under a static market structure.  Here, we are interested in the mechanisms that can handle continuous data streams, hence real-time data markets. In this direction, the authors in \cite{agarwal2019marketplace} present a real-time data market for buying and selling training data and propose a mechanism to fairly compensate the data providers. However, the compensation in their mechanism is computed offline. In the presence of continuous data streams and combinatorial complexity of data valuation, offline solutions cannot be executed in the time scales that match the dynamics of the underlying process.  {Therefore, in this paper, we adopt a game-theoretic approach to design online market mechanisms for real-time markets operating at fast time scales. We present these mechanisms in a general form that is applicable to several domains. For the data markets in \cite{agarwal2019marketplace}, the online formulation enables us to better remunerate the market players under continuous data streams. Similarly, for  the real-time P2P market of \cite{guo2021online}, we can employ online market mechanisms grounded in coalitional game theory, which offers mathematical tools for analysing the interaction of self-interested agents and provides guarantees of \textit{fairness} or stability on the remuneration criteria. From an economic perspective, these properties are highly desirable for a payoff distribution mechanism.}




In this paper, we focus on a particular class of coalitional games, namely transferable utility (TU) coalitional game, which consists of a set of agents $\mathcal{I}$ and a value function $v$ that assigns a value $v(S)$ to each possible coalition of agents $S \subset \mathcal{I}$. Collectively, a TU coalitional game is represented by a pair $(\mathcal{I},v)$ \cite{myerson2013game}. Multi-agent decision-making problems modeled by coalitional games arise in many application areas, such as energy systems \cite{han2018constructing}, \cite{chakraborty2018sharing} and communication networks \cite{Saad2009}. In particular, we study markets modeled as coalitional games. Coalitional game theory studies the mechanism of the distribution of the value generated by cooperation to respective agents. Two key solution concepts that undertake the task of value distribution (payoff) are the Shapley value and the core. The Shapley value addresses the fairness aspect, which implies that the payoff for an agent should reflect its impact on the game. This property is ensured by the axiomatic characterization of fairness \cite{shapley1953value}. The core payoff ensures that no agent has any incentive to defect the coalition and thus addresses stability.

We consider the problems of evaluating both fair and stable payoff allocations, i.e., the Shapley value and the core payoff respectively, under a dynamic coalitional game setting. Essentially, our work lies at the intersection of time-varying optimization and dynamic coalitional games. In the direction of the former, algorithms proposed in the literature \cite{Simonetto2020}, \cite{Simonetto2015} track trajectories of the optimizers of the time-varying optimization problems up to asymptotic error bounds, under the assumption of strong convexity.
The problem of payoff allocation in dynamic coalitional games has also been studied in the literature. Among others, the authors in \cite{lehrer2013core} characterize the core allocations when the coalitional values vary over time and are dependent on previous events. In \cite{bauso2009robust}, Bauso and Timmer propose payoff allocation rules for a dynamic game where the coalitional value fluctuates within a bounded polyhedron while the average value of each coalition over time is known. As we are seeking to design iterative algorithms, a closer work is \cite{Nedic2013} by Nedich and Bauso. The paper considers a core payoff allocation in a sequence of games where the intersection of all the corresponding cores is non-empty. Further generalization of their work is presented by the authors in \cite{raja2021payoff} under the same assumption on the core sets.  {However, in the context of real-time markets it is not reasonable to assume that the coalitional values evolve only within a particular set or that we have knowledge about the average coalitional value over time. Thus, the assumptions made on the non-empty intersection of the solution sets in the works mentioned above make their algorithms inapplicable to real-time markets. Considering these short-comings, in this paper, we drop the assumptions on the knowledge of average coalitional values as well as of non-empty intersection of solution sets to formulate coalitional games in an online paradigm and in turn propose solutions for real-time market setups. 
}

 {A typical problem of real-time markets modelled as a coalitional game is the exponential computational complexity of an equilibrium solution which usually makes exactly evaluating the core and Shapley value impractical. Therefore, we introduce online distributed payoff allocation algorithms that instead of evaluating at each time instant the exact solution, track the solutions of the continuously-varying coalitional games up to an asymptotic error bound. Among all energy-related markets, here we focus on the advanced real-time markets that are operating at a high frequency, where the time interval between the opening and the clearing of the market is not enough to compute a coalitional solution in an offline manner. We note that online mechanisms are instead not necessary or suitable for traditional centralised wholesale markets. }


Before listing our contributions, let us further motivate our setting through an example. Note that, here, we present our example in a general setting to show the extent of our contribution. Later, we simulate the energy-related market as a specific case of this motivational setting. 

\begin{figure}
\centering
\includegraphics[width=0.75\columnwidth]{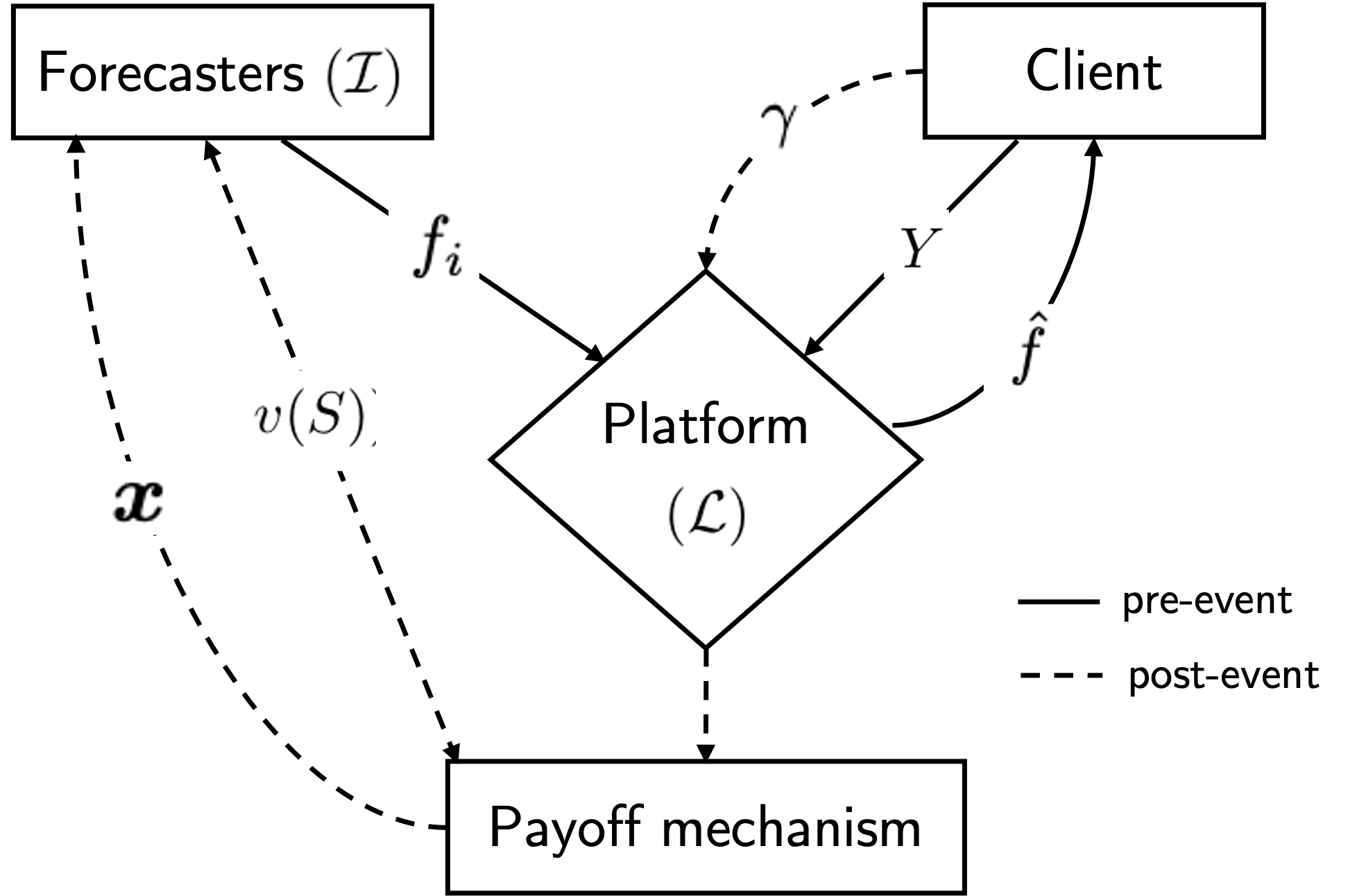}
\caption{Overview of an online data valuation scheme in the context of a collaborative forecasting market.}
\label{fig: prediction market}
\vspace{-0.5cm}
\end{figure}

\textit{Motivational example}: Let us consider an online forecast valuation scheme, inspired by \cite{kilgour2004elicitation} and \cite{raja2022market}, for pooling the information and expertise held by different owners and generating a combined forecast. First, let us introduce the forecasting markets that are designed to predict an event e.g. renewable energy generation \cite{shamsi2021prediction}. Generally, in such markets, the market participants (forecasters) sell predictions in the form of a probability distribution; then the true outcome of the event is observed and the market pays each expert based on the \textit{quality} of their predictions. 
Let there be a central platform $\mathcal{L}$ designed for a prediction task, e.g. to predict wind energy generation. Consider a set of $N$ forecasters, $\mathcal{I}$, that have expertise in making such predictions. To generate accurate predictions, the forecasters take into account various factors, e.g. wind speed and overall weather conditions affecting wind energy generation. Each provider $i \in \mathcal{I}$ posts a bid to the forecasting market. To achieve the forecast valuation, the following steps are performed:
\begin{itemize}
    \item A client posts a prediction task $Y$ to the central platform;
    \item Each forecaster $i \in \mathcal{I}$ posts their  prediction $f_i$ of the announced task;
    \item The platform combines these forecasts using a pooling method \cite{wallis2005combining} and the resulting aggregate forecast $\hat{f}$ is delivered to the client;
    \item After the event occurs, the client announces a reward $\gamma$ corresponding to the improvement that they achieved in decision making. Then, the quality of posted predictions is evaluated and the reward is distributed fairly among the forecasters as a payoff $\boldsymbol{x}$. 
\end{itemize}
In our setting, we consider high-frequency events with fast dynamics which thus requires an online forecast valuation scheme.  
This process of eliciting a combined forecast, i.e., collaborative forecasting results in an online coalitional game among the forecasters, represented by a triplet $(\mathcal{I}, \mathcal{L}, v^k)$. The setup of real-time valuation  results in a time-varying value function $v^k$, where $v^k(S)$ represents the utility of a client attained by a cooperative forecast of coalition $S \subset \mathcal{I}$. After the occurrence of the event, forecasters negotiate their share of the resulting value according to a criterion that acknowledges their individual contributions in predicting that event. In Fig. \ref{fig: prediction market}, we present a collaborative forecasting scheme with an online payoff distribution mechanism.
In the literature, the Shapley value is utilized for payoff allocation in an offline setting for similar markets \cite{jia2019efficient}, \cite{Agarwal2019} as it fulfills the key criterion of a fair forecast valuation scheme. For further details on the criterion, we refer to \cite{ghorbani2019data}. In this paper, we design online algorithms for the most widely used payoff distribution methods in coalitional games, namely the core and the Shapley value.

\textit{Contribution}: 
\begin{itemize}
\item We introduce the concept of online tracking of solutions (Shapley value and the core) in the context of coalitional game theory;
\item We develop a novel distributed online payoff allocation algorithm to track the Shapley value up to an asymptotic error bound. We also present the static version of the algorithm which converges to the Shapley value exactly (Section \ref{sec: payoff allocation});
\item {We relax the assumption on the core sets of the sequence of coalitional games in \cite{raja2021payoff}}  and present an online algorithm to track the payoff allocation in a neighborhood of the core. We show that the proposed algorithm is asymptotically consistent, i.e., converges to the core payoff exactly in the absence of dynamics (Section \ref{subsec: Online tracking of core allocation});
\item {We introduce an operator theoretic analysis for the design of online algorithms in the domain of dynamic coalitional games, which allows us to generalize existing results.} 
\end{itemize}
We note that, even though we focus on the mechanism design of energy-related real-time markets, our solutions can be applied to other applications of cooperative game theory as well. For instance, a community-based energy storage optimisation presented in \cite{han2018constructing} can be addressed in an online fashion to mitigate the effects of uncertainty in load and RES generation. 
Similarly, real-time fair pricing can be achieved for a ride-hailing service proposed in \cite{shao2021fair}.


\textit{Notation}: $\mathbb{R}$ and $\mathbb{N}$ denote the set of real and natural numbers, respectively. Given a mapping $M: \mathbb{R}^n \rightarrow \mathbb{R}^n, \mathrm{fix}(M):= \{x \in \mathbb{R}^n \mid x = M(x)\} $ denotes the set of its fixed points. $\text{Id}$ denotes the identity operator. For a closed set \(C \subseteq \mathbb{R}^{n},\) the mapping $\mathrm{proj}_C$: \(\mathbb{R}^{n} \rightarrow C\) denotes the projection onto \(C,\) i.e., \(\operatorname{proj}_C(x)=\) \(\arg \min _{y \in C}\|y-x\| .\) For a set $S$ the power set is denoted by $2^S$. \(A \otimes B\) denotes the Kronecker product between the matrices \(A\) and \(B .\) $I_N$ denotes an identity matrix of dimension $N \times N$. {For $x_{1}, \ldots, x_{N} \in \mathbb{R}^{n},$ $\mathrm{col}(\left(x_{i}\right)_{i \in(1, \ldots, N)}):=\left[x_{1}^{\top}, \ldots, x_{N}^{\top}\right]^{\top}.$}  $\mathrm{dist}(x,C)$ denotes the distance of $x$ from a closed set \(C \subseteq \mathbb{R}^{n},\) i.e., $\mathrm{dist}(x,C):= \mathrm{inf}_{y \in C} \|y-x\|$. {For a closed set \(C \subseteq \mathbb{R}^{n}\) and $N \in \mathbb{N}, C^N:=\prod_{i=1}^{N} C_{i}$}.

\textit{Operator-theoretic definitions}:
A mapping $T : \mathbb{R}^n \rightarrow \mathbb{R}^n$ is contraction, if $ \|T(x) - T(y)\| < \|x - y\|,$
for all $x,y \in \mathbb{R}^n$.

\section{Background on Coalitional Games}\label{sec:Mathematical background }
Let us provide the necessary mathematical  background  on coalitional game theory in a dynamic context and describe an online payoff distribution process. 
 \begin{definition}[Dynamic coalitional game {(\cite{Nedic2013}, Sec. II-A)}]\label{def: instantaneous TU coalitional game} 
Let $\mathcal{I} = \{1, \ldots, N\}$ be a set of agents. For each time $k \in \mathbb{N},$ an instantaneous coalitional game is a pair $\mathcal{G}^k = (\mathcal{I}, v^{k})$ where $v^k: 2^\mathcal{I} \to \mathbb{R}$ is a value function that assigns a real value, $v^k(S)$, to each coalition $S \subseteq \mathcal{I}$. A dynamic coalitional game is a sequence of instantaneous games, i.e., $\mathcal{G} = (\mathcal{I}, (v^{k})_{k \in \mathbb{N}})$.$\hfill \square$
 \end{definition}

\noindent For each instantaneous game, the value generated by a coalition has to be distributed among its members as a payoff $x_i^k$, which represents a share of agent $i$ of the value $v^k  (\mathcal{I})$.
The goal is to find a payoff vector with desirable properties like \textit{stability} and \textit{fairness}. The solution concept that relates to the \textit{stability} of a grand coalition, i.e., a coalition of all agents is the core whereas, the \textit{fairness} axioms are satisfied by the Shapley value \cite{shapley1953value}. Note that the Shapley value does not necessarily belong to the core. 
Let us first define a set of stable payoff vectors for each player, i.e., a \textit{bounding set} and then define the core, which is an intersection of all bounding sets. 
\begin{definition}[Bounding set (\cite{Nedic2013}, Sec. II-B )]\label{def: bounding set}
For an instantaneous game $\mathcal{G}^{k}=(\mathcal{I}, v^{k}), k \in \mathbb{N}$, the set
\begin{equation} \label{eq: bounding set}
\begin{array}{ll}
\mathcal{X}_i(v^{k}) :=    \bigg\{& x \in \mathbb{R}^N \mid \sum_{j \in \mathcal{I}} x_j = v^k (\mathcal{I}),\\
  &  \sum_{j \in S} x_j \geq v^k  (S), \forall S \subset \mathcal{I} \text{ s.t. } i \in S \bigg\}
\end{array}
\end{equation}
denotes the bounding set of an agent $i \in S$. $\hfill \square$
\end{definition}

For an instantaneous coalitional game the core is defined as follows.
\begin{definition} [Instantaneous core]\label{def: inst core}
The core $\mathcal{C}$ of an instantaneous coalitional game $\mathcal{G}^{k} = (\mathcal{I},v^{k}), k \in \mathbb{N}$, is the following set of payoff vectors:
\begin{equation} \label{core}
\begin{array}{lll}
  \mathcal{C}(v^{k}) &:=  \bigg\{ x \in \mathbb{R}^N \mid  \sum_{i \in \mathcal{I}} x_i = v^k (\mathcal{I}),\\
  &\qquad \qquad \qquad \;  \sum_{i \in S} x_i \geq v^k  (S), \forall S \subseteq \mathcal{I}  \bigg\},
\end{array}
\end{equation}
 $\hfill \square$
\end{definition}
In the sequel we deal with the core solution which is assumed to be non-empty.
\begin{assum}\label{asm: nonempty core}
The core of each instantaneous game $(\mathcal{I}, v^k)$, is non-empty, i.e., $\mathcal{C}(v^k) \neq \varnothing \text{ for all } k \in \mathbb{N}$.  $\hfill \square$
\end{assum}
The core set does not comply with the notion of fairness in fact different core allocations treat agents differently. 
The unique payoff allocation that satisfies \textit{fairness} axioms (see \cite{shapley1953value}) is known as the Shapley value. 
\begin{definition} [Shapley value]\label{def: shapley value}
For a coalitional game $\mathcal{G}^{k} = (\mathcal{I},v^{k})$, let $\Pi$ be the set of all ($N!$) permutations of the grand coalition $\mathcal{I}$ and, for an ordering of agents $\sigma \in \Pi$, let $\mathcal{P}_i^{\sigma}$ be the set of predecessors of $i$ in $\sigma$ with $\mathcal{P}_1^{\sigma} = \varnothing$. Then, for every player $i \in \mathcal{I}$ the Shapley value $\boldsymbol{\phi}(v)$ assigns the payoff $\phi_{i}(v)$ given by:
\begin{equation}\label{eq: shapley value}
\phi_{i}(v^k)=\frac{1}{N!} \sum_{\sigma \in \Pi}(v^k(\mathcal{P}_i^{\sigma} \cup \{i\})-v^k(\mathcal{P}_i^{\sigma})).
\end{equation}
 $\hfill \square$
\end{definition}
Here, we refer to the term  $(v^k(\mathcal{P}_i^{\sigma} \cup \{i\})-v^k(\mathcal{P}_i^{\sigma}))$ in (\ref{eq: shapley value}) as the incremental marginal contribution which shows the value added by an agent $i$ when it joins the coalition. \\
We note from (\ref{eq: shapley value}) that to evaluate its Shapley payoff an agent needs to know the value of all possible coalitions, which is impractical for many real-world applications and renders distributed computation useless. For the purpose of designing a distributed algorithm, we identify the orderings $\sigma \in \Pi$ for which an agent $i$ can evaluate the incremental marginal contributions of all the agents with only the knowledge of the coalitional values of its own coalitions. These orderings are the ones in which $i$ joins the coalition at the first position. To clarify further, we present the following example.
\begin{example}\label{exm: marginal contribution}
Let us consider a three player coalitional game $(\{a,b,c\},v)$. Here,  agent $a$ can compute the incremental marginal contributions for ordering $(a,b,c)\text{ and } (a,c,b)$ as $(v(\{a\},v(\{a,b\} - v(\{a\}, v(\{a,b,c\}-v(\{a,b\} ) \text{ and } (v(\{a\},v(\{a,c\} - v(\{a\}, v(\{a,b,c\}-v(\{a,c\} ))$, respectively by knowing the values of its own coalitions only. However, for the ordering $(b,c,a)$ the incremental marginal contributions are $(v(\{b\},v(\{b,c\} - v(\{b\}, v(\{a,b,c\}-v(\{b,c\} )$ and to evaluate them, agent $a$ requires the knowledge of $v(\{b,c\})$ which is unreasonable as the coalition $(b,c)$ is not its coalition.   $\hfill \square$
\end{example}
To exploit the observation from Example \ref{exm: marginal contribution}, in the sequel, we define the marginal contribution vector $\boldsymbol{\hat{m}}_i$ that is the average of incremental marginal contribution vectors corresponding to those orderings for which an agent $i$ can evaluate with minimal information.
\begin{definition} [Marginal contribution vector]\label{def: marginal contribution}
Let $\pi_i \subset \Pi$  be the set of permutations of the grand coalition $\mathcal{I}$ in which an agent $i$ occupies the first position. For each ordering $\sigma \in \Pi$, let $\boldsymbol{m}_\sigma \in \mathbb{R}^N$ be a vector of incremental marginal contributions with $j\text{th}$ element $m^\sigma_j = v^k(\mathcal{P}_j^{\sigma} \cup \{j\})-v^k(\mathcal{P}_j^{\sigma})$.   Then, for every agent $i \in \mathcal{I}$, the marginal contribution vector is
\begin{equation}\label{eq: marginal vector}
\boldsymbol{\hat{m}}_i=\frac{1}{(N-1)!} \sum_{\sigma \in \pi_i}\boldsymbol{m}_\sigma.
\end{equation}
 $\hfill \square$
\end{definition}
Now, the Shapley value, in terms of marginal contribution vectors, becomes $\boldsymbol{\phi}(v) = \frac{1}{N} \sum_{i \in \mathcal{I}}\boldsymbol{\hat{m}}_i$.\\
\noindent Next, we note that for a dynamic coalitional game the solution also varies with time and that the complexity of both the solutions, i.e., the core in (\ref{core}) and the Shapley value in (\ref{eq: shapley value}) grows exponentially with the number of agents. Therefore, guaranteeing convergence to a solution payoff vector for each instantaneous game  is not necessarily possible, especially in highly dynamic settings, e.g. real-time applications, where computational and communication bottlenecks can hinder the exact tracking of a solution trajectory. Therefore, in the sequel, we propose a distributed online algorithm to track the Shapley value and provide bounds on the asymptotic error, defined as the ``distance" between the evaluated payoff vector and the Shapley value. Furthermore, we also design a distributed online algorithm that provides a bound on the asymptotic error for tracking the core set. \\  
 \noindent In a distributed online payoff allocation method an agent $i$ proposes a payoff distribution $ \boldsymbol{x}_i \in \mathbb{R}^N$ according to a criteria. The allocation procedure aspires to reach a mutually agreed payoff (consensus) in the core.
\begin{definition} [Consensus set]\label{def: consensus set}
The consensus set $\mathcal{A} \subset \mathbb{R}^{N^{2}}$ is defined as:
\begin{equation}\label{eq: consensus}
\mathcal{A} := \{\mathrm{col}(\boldsymbol{x}_1, \ldots, \boldsymbol{x}_N) \in \mathbb{R}^{N^{2}} \mid \boldsymbol{x}_i = \boldsymbol{x}_j, \forall i,j \in \mathcal{I}\}. 
\end{equation} $\hfill \square$
\end{definition}
In the sequel, first, we consider the problem of computing a trajectory of payoff vectors that  converges to the Shapley value up to a bounded error, i.e., $\displaystyle \lim_{k \to \infty} \sup \| \boldsymbol{x}^k - \boldsymbol{\Phi}^k\|$ is small. Then, we address the problem of tracking the core set such that $ \displaystyle\lim_{k \to \infty}\sup \mathrm{dist}(\boldsymbol{x}^k,\mathcal{A} \cap \mathcal{C}(v^k))$ is small. 

\vspace{-2.5mm}
\section{Distributed Online Payoff Allocation}\label{sec: payoff allocation}
In this section, we propose a payoff distribution in the context of online coalitional games, where the value function $v$ varies with time $k$ on a fast scale hence, the exact computation of the solution for each instantaneous game is not computationally achievable. Therefore, our goal is to design a distributed algorithm to compute a payoff trajectory that tracks a solution reasonably well. We remark that we analyse the most conservative case where agents evaluate one iteration per sample $v^k$. The tracking performance can be improved with multiple iterations per sample, depending on the lead time of a market. \\
Let a set of agents $\mathcal{I} = \{1, \ldots, N\}$ synchronously propose a distribution of utility at each discrete time step $k \in \mathbb{N}$, i.e., each agent $i \in \mathcal{I}$ proposes a payoff distribution ${ \boldsymbol{x}}_i^{k} \in \mathbb{R}^N$, where the $j$th element denotes the share of agent $j$ proposed by agent $i$ at step $k \in \mathbb{N}$.\\
Let the agents communicate over a time-varying network represented by a graph $G^{k}=(\mathcal{I}, \mathcal{E}^{k}) $, where $(j, i) \in \mathcal{E}^{k}$ means that there is an active link between the agents $i$ and $j$ at iteration $k$ and they are then referred as neighbours. Therefore, the set of neighbors of agent $i$ at iteration $k$ is defined as \(\mathcal{N}_{i}^k:=\left\{j \in \mathcal{I} |(i, j) \in \mathcal{E}^{k}\right\}\). We assume that at each iteration $k$ the communication graph is connected.
The edges in the communication graph $G^{k}$ are weighted using an adjacency matrix $W^{k} = [w_{i,j}^k]$, whose element $w_{i,j}^k$ represents the weight assigned by agent $i$ to the payoff distribution proposed by agent $j$, ${ \boldsymbol{x}}_j^{k}$. Note that, for some $j$, $w_{i,j}^k = 0$ implies that $j \notin \mathcal{N}_{i}^k $ hence, the state of agent $i$ is independent from that of agent $j$. We assume the adjacency matrix to be doubly stochastic with positive diagonal elements, as assumed in \cite[Assumption 3.3]{nedic2017achieving}, \cite[Assumptions 2, 3]{nedic2010constrained}. 
\begin{assum}[Stochastic adjacency matrix]\label{asm: graph}
 For all $k \geq 0$, the adjacency matrix $W^{k} = [w_{i,j}^k]$ of the communication graph $G^k$ satisfies following conditions:
    \begin{enumerate}
        \item It is symmetric and doubly stochastic, i.e., \(\sum_{j=1}^{N} w_{i,j}=\sum_{i=1}^{N} w_{i,j}=1\);
        \item its diagonal elements are strictly positive, i.e., $w_{i,i}^k > 0, \forall i \in \mathcal{I}$;
        \item $\exists$ $\gamma > 0$ such that $w_{i,j}^k \geq \gamma$ whenever $w_{i,j}^k > 0$. $\hfill \square$
    \end{enumerate} 
\end{assum}
{Assumption \ref{asm: graph} ensure that the agents communicate sufficiently often to each other and have sufficient influence on the resulting allocation.} 
Finally, we propose distributed discrete-time algorithms of the form:
$$
\boldsymbol{x}_i^{k+1} = M_i^k (\boldsymbol{x}_i^{k}),
$$
where $\boldsymbol{x}_i^{k} \in \mathbb{R}^N$ is agent $i$'s estimate of the payoff allocation of all the agents and   $M_i^k $ is a time-varying update operator. We can write the above iteration for all agents in collective compact form:
\begin{equation}\label{main_it_compact}
\boldsymbol{x}^{k+1} = \boldsymbol{M}^k (\boldsymbol{x}^{k}),
\end{equation}
where $\boldsymbol{M}^k (\boldsymbol{x}):= \mathrm{col}(M_1^{k}(\boldsymbol{x}_1), \ldots,  M_N^{k}(\boldsymbol{x}_N)) $. Next, we assume a bound on the time variation of the fixed-point of the time-varying operators $\boldsymbol{M}^k$ in (\ref{main_it_compact}).
 \begin{assum}[Bounded time variations]\label{asm: Bounded time variations}
 Let $(\boldsymbol{M}^{k})_{k \in \mathbb{N}}$ be the sequence of operators in (\ref{main_it_compact}). The distance between the fixed-points of two consecutive operators is bounded, i.e., $\sup_{k \in \mathbb{N}} \sup_{(\bar{\boldsymbol{x}}^k, \bar{\boldsymbol{x}}^{k+1}) \in \mathrm{fix}(\boldsymbol{M}^{k}) \times \mathrm{fix}(\boldsymbol{M}^{k+1})} $ $||\bar{\boldsymbol{x}}^{k-1} - \bar{\boldsymbol{x}}^k|| \leq \delta, \text{ for some } \delta > 0$. $\hfill \square$
 \end{assum}
 
 We note that Assumption \ref{asm: Bounded time variations} bounds the time variations of the fixed point sets of the time-varying operators, rather than the Euclidean distance between the optimal points at consecutive times, i.e., $||\bar{\boldsymbol{x}}^{k-1} - \bar{\boldsymbol{x}}^k|| \leq \delta$, which is standard in the time-varying optimization  \cite[Assumption 1]{Simonetto2020}, \cite[Theorem 1]{simonetto2015prediction}. Next, we present an online payoff allocation algorithm  where we design the operator $M_i^k$ with the Shapley payoff as its fixed-point set, i.e.,  $\mathrm{fix}(M_i^k) = \boldsymbol{\phi}(v^k)$. We note that in the context of Shapley payoff distribution, Assumption \ref{asm: Bounded time variations} relates to the dynamics of the coalitional game and implies a bounded variation of the Shapley value from one time step to the next. 
 
\subsection{Online tracking of the Shapley allocation}\label{subsec: Shapley allocation}
Let us formulate the distributed tracking of the Shapley value via time-varying operators and provide convergence results. 
The problem of computing the Shapley value for a static coalitional game can be formulated as an unconstrained convex optimization problem with the objective of achieving a consensus on the Shapley value, i.e.,  
\begin{equation}\label{eq: shapley optimization}
 \underset{\boldsymbol{x}}{\operatorname{min }} \frac{1}{2} \sum_{i \in \mathcal{I}}\left\|\boldsymbol{x}-\hat{m}_{i}\right\|^{2}, 
\end{equation}
where $\hat{m}_{i}$ is a marginal contribution vector as in (\ref{eq: marginal vector}). Here, we consider dynamic coalitional games executed on time-varying networks and design an algorithm in a distributed paradigm, thus the marginal contribution vector is also time-varying. Each agent $i$ minimizes a local objective function $f_i^k = \frac{1}{2} \left\|\boldsymbol{x}_i-\hat{m}^k_{i}\right\|^{2}$. To solve the resulting optimization problem, an agent $i$ can adopt a gradient based algorithm. Let $\textstyle \boldsymbol{y}_i^k := \textstyle \sum_{j =1}^N w_{i,j} \boldsymbol{x}_j$, then the state update is given as $$\boldsymbol{x}_i^{k+1} =  \boldsymbol{y}_i^k-\alpha \nabla f_{i}^k(\boldsymbol{y}_i^k).$$ 
In operator-theoretic terms, we can define an operator $\boldsymbol{M}^k$ in (\ref{main_it_compact}) as a composition of a gradient step operator and a consensus operator, i.e.,   $\boldsymbol{M}^k = (\mathrm{Id}-\alpha \nabla f^k)\circ \boldsymbol{W}^k $ where $\boldsymbol{W}^k := W^{k} \otimes I_N $ represents an adjacency matrix.
We note that for a strongly convex function $f$ the operator $\boldsymbol{M}$ is a contraction mapping, a fact we use later to prove the convergence of the proposed algorithm. 

 \begin{assum}[Contractions]\label{asm: contraction}
 For all $k \in \mathbb{N}$, the operator $\boldsymbol{M}^k$ in (\ref{main_it_compact}) is such that $\boldsymbol{M}^k \in \mathcal{M}$, where $\mathcal{M}$ is a family of contraction operators with contraction factor $L_k \in (0,1)$. 
 \end{assum} $\hfill \square$\\
Finally, the compact and simplified iteration takes the following form:
\begin{equation}\label{main_it_shapley}
\boldsymbol{x}^{k+1} = (1-\alpha)\boldsymbol{W}^k\boldsymbol{x}^{k} + \alpha \boldsymbol{\hat{m}}^k .
\end{equation}
The authors in \cite{yuan2016convergence} present an iteration based on the static version of the operator $\boldsymbol{M}$, i.e., $\boldsymbol{z}^+ = \boldsymbol{M} \boldsymbol{z} $ and show an inexact convergence which achieves an asymptotic error bound $O(\alpha)$ with respect to the consensus optimizer $\boldsymbol{\bar{z}}$. In our setting, $\boldsymbol{\bar{z}}$ refers to the Shapley value $\boldsymbol{\phi}$. Here we derive a bound for an online setting in terms of $O(\alpha)$-neighborhood as defined in \cite[Lemma 1]{yuan2016convergence} under the same conditions on the step size $\alpha$.
We note that the solution in the context of online coalitional games means convergence of the payoff allocation trajectory to a neighbourhood of the time-varying Shapley value, as shown by the following convergence result for (\ref{main_it_shapley}).

\begin{theorem}[Convergence of online Shapley allocation]\label{thm: online Shapley allocation}
Let Assumptions \ref{asm: graph}$-$ \ref{asm: contraction} hold. Starting from any $\boldsymbol{x}^0 \in \mathbb{R}^{N^2}$, the error norm $|| \boldsymbol{x}^{k} - \boldsymbol{\Phi}^{k} ||$ generated by the iteration in (\ref{main_it_shapley}) satisfies the following bound:
$$
\|\boldsymbol{x}^{k}-\boldsymbol{\Phi}^{k}\| \leqslant \hat{L}_{k}\|\boldsymbol{x}^{0}-\boldsymbol{\Phi}^{0} \|+\frac{1-(\bar{L}_{k})^{k-1}}{1-\bar{L}_{k}} \delta + O(\alpha),
$$
where $\hat{L}_{k}:=\prod_{i=1}^{k-1}L_{i}, \bar{L}_{k}=\max _{k} L_{k}$ and $\boldsymbol{\Phi}^k:= \boldsymbol{\phi}^k \otimes \boldsymbol{1}_N$ is the Shapley value in (\ref{eq: shapley value}) where $O(\alpha)$ is as in \cite[Lemma 1]{yuan2016convergence}. Therefore, we have that $\lim_{k \to \infty}$  $\textstyle
\|\boldsymbol{x}^{k}-\boldsymbol{\Phi}^{k}\| \leqslant \frac{1}{1-\bar{L}_{k}} \delta +O(\alpha)
$. 
\end{theorem} $\hfill \square$\\
The result of Theorem \ref{thm: online Shapley allocation} asserts that the sequence  $(\boldsymbol{x}^k)_{k \in \mathbb{N}}$ tracks the trajectory of the Shapley value up to a bound that linearly depends on the parameter $\delta$, which comes from Assumption $\ref{asm: Bounded time variations}$ and relates to the time variability of the Shapley allocation of a dynamic coalitional game in Definition \ref{def: instantaneous TU coalitional game}. We provide the proof of Theorem \ref{thm: online Shapley allocation} in Appendix.

We note that if the coalitional game is static then by using the setting in \cite[Theorem 1]{yamada2005hybrid} we can design a distributed  algorithm that converges to the Shapley allocation.  Let us present a corollary for the static case.
\begin{corollary}[Convergence to Shapley allocation]\label{cor: payoff allocation contraction}
Let Assumptions \ref{asm: graph} and \ref{asm: contraction} hold. Let Assumption \ref{asm: Bounded time variations} hold with $\delta=0$. Then, starting from any $\boldsymbol{x}^0 \in \mathbb{R}^{N^2}$, the sequence \((\boldsymbol{x}^{k})_{k=0}^{\infty}\) generated by the iteration $$\boldsymbol{x}^{k+1} = (1-\alpha_k)\boldsymbol{W}^k\boldsymbol{x}^{k} + \alpha_k \boldsymbol{\hat{m}}^k,$$ converges to the Shapley value in (\ref{eq: shapley value}), i.e., $\boldsymbol{x}^k \to \boldsymbol{\Phi}, $ where $\left(\alpha_{k}\right)_{k \in \mathbb{N}} \in (0,1)$ such that $\alpha_{k} \rightarrow 0, \sum_{k \in \mathbb{N}} \alpha_{k}=+\infty$, $\sum_{k \in \mathbb{N}}\left|\alpha_{k+1}-\alpha_{k}\right|<+\infty$. $\hfill \square$
\end{corollary}
\textit{Discussion:}
The solutions offered by coalitional game theory have interesting mathematical properties, but their computational complexity poses a challenge to their utilization in real-world applications. As the evaluation of the Shapley value requires the computation of the value of all possible permutations of the set of agents, the computational time increases exponentially with the number of agents. This challenge makes it impractical to utilize the Shapley payoff allocation in almost real-time. In this direction, the distributed structure of proposed algorithm in (\ref{eq: shapley value}) mitigates the problem of high computational times by logically distributing the computational burden among the agents. Furthermore, it democratizes the negotiation process by autonomizing the decision making of agents, which is an important feature of liberal markets.

For coalitional games, the payoff allocated via the Shapley value guarantees fairness. However, it does not ensure the stability of a grand coalition $\mathcal{I}$, i.e., the Shapley value does not necessarily belong to the core in (\ref{core}). As a consequence, if a coalition structure is not encouraged externally, then the Shapley payoff might not provide an adequate incentive for agents to join a coalitional game. Therefore, it is highly desirable to design a distributed algorithm for an online tracking  of the core in dynamic coalitional games. 
\subsection{Online tracking of a core allocation}\label{subsec: Online tracking of core allocation}
Let us now turn our attention towards the problem of tracking the core solution in (\ref{core}) for online coalitional games. As the core is a set which in dynamic game setting varies with time, the problem takes the form of distributively tracking a time-varying set. For an agent $i$, the problem of tracking the core set $\mathcal{C}(v^k)$ can be formulated as an unconstrained time-varying convex optimization problem with objective of minimizing the distance of agent's payoff allocation estimate from its bounding set in (\ref{eq: bounding set}). Mathematically, each agent $i$  has an objective function  $ \textstyle f_i^k := \frac{1}{2} \|\hat{\boldsymbol{x}}_i^k-\mathrm{proj}_{\mathcal{X}_i(v^k)}(\hat{\boldsymbol{x}}_i^k)\|^{2}  + \frac{\gamma}{2} \| \hat{\boldsymbol{x}}_i^k - \boldsymbol{x}_i^{k-1}\|^{2}$ with $\gamma > 0$. Thus, the optimization problem takes the following form:
 
\begin{equation}\label{optimization_prob}
\left\{
\begin{array}{l}
 \underset{\boldsymbol{x}_i}{\text{min }} \frac{1}{2} \|\hat{\boldsymbol{x}}_i^k-\mathrm{proj}_{\mathcal{X}_i(v^k)}(\hat{\boldsymbol{x}}_i^k)\|^{2} + \frac{\gamma}{2} \| \hat{\boldsymbol{x}}_i^k - \boldsymbol{x}_i^{k-1}\|^{2} \\
\text{s.t. } \hat{\boldsymbol{x}}_i^k = \sum_{j=1}^{N} w_{i,j}^k \boldsymbol{x}_{j}^{k}
\end{array}
\right.
\end{equation}
The optimization problem in (\ref{optimization_prob}) can be solved by using an iteration based on the forward operator $\mathrm{Id}- \alpha \nabla f_{i}$ which is a contraction mapping for a strongly convex and a strongly smooth function $f_i$. 
In our setup, for each time step $k \in \mathbb{N}$, an agent $i$ updates its state as
\begin{equation}\label{main_it_core}
\boldsymbol{x}_i^{k+1} = (1-\alpha - \alpha \gamma) \hat{\boldsymbol{x}}_i^k + \alpha \mathrm{proj}_{\mathcal{X}_i(v^k)}(\hat{\boldsymbol{x}}_i^k) + \alpha \gamma \boldsymbol{x}_i^{k-1},
\end{equation}
where $\mathcal{X}_i$ is a bounding set in (\ref{eq: bounding set}). In a stacked vector notation the forward operator applied on (\ref{optimization_prob}) gets composed with the consensus operator, i.e.,  $(\mathrm{Id}- \nabla f) \circ \boldsymbol{W} (\cdot)$.  Let us further generalize the iteration in (\ref{main_it_core}) by replacing the projection operator, $\mathrm{proj}(\cdot)$, with contractions in Assumption \ref{asm: contraction}. This generalization enables the agents to choose any contraction operator $T_i^k$ for evaluating a payoff $\boldsymbol{x}_i^k$. For consistency we require the fixed-point set of $T_i^{k}$ to be the bounding set in (\ref{eq: bounding set}), i.e., $\mathrm{fix}(T_i^k) = \mathcal{X}_i(v^k)$. Consequently, $ \mathrm{fix}(\boldsymbol{T}^k) = \bigcap_{i=1}^N \mathcal{X}_i(v^k) = \mathcal{C}(v^{k})$, the instantaneous core set. The contraction property allows us to prove the convergence of the state $\boldsymbol{x}^k$ to the set $\mathcal{A}\cap \mathcal{C}^N$ up to a specified error bound. Specifically, we propose the following online allocation protocol:
 \begin{equation}\label{main_it_core_general}
    { \boldsymbol{x}}^{k+1} = \boldsymbol{T}^k (\boldsymbol{W}^k { \boldsymbol{x}}^{k}),
    \end{equation}
where the operator $\boldsymbol{M}^k:= \boldsymbol{T}^k (\boldsymbol{W}^k(\cdot))$ as in iteration (\ref{main_it_compact}) is a sequence of time-varying contraction operators corresponding to the time-varying core set being tracked via a time-varying communication network.  {This formulation of online tracking in terms of operators allows us to use the existing results from operator theory and to generalize the algorithms in \cite{Nedic2013} and \cite{raja2021payoff} by dropping their assumption that the intersection of time-varying cores is non-empty. Furthermore, the operator theoretic analysis allows us to keep our proofs brief and elegant.}
 Next, we formalize the convergence result of online tracking of the core allocation.
  \begin{theorem}[Online core payoff allocation]\label{thm: payoff allocation paracontraction}
 Let Assumptions \ref{asm: nonempty core}$-$\ref{asm: contraction} hold. Then, {starting from any $\boldsymbol{x}^0 \in \mathbb{R}^{N^2}$,} the error norm $|| \boldsymbol{x}^{k} - \bar{\boldsymbol{x}}^{k} ||$ generated by the iteration in (\ref{main_it_compact}) satisfies the following bound: 
$$
\|\boldsymbol{x}^{k}-\bar{\boldsymbol{x}}^{k}\| \leqslant \hat{L}_{k}\|\boldsymbol{x}_{0}-\bar{\boldsymbol{x}}_{0} \|+\frac{1-(\bar{L}_{k})^{k-1}}{1-\bar{L}_{k}} \delta,
$$
where $
\hat{L}_{k}=\prod_{i=1}^{k-1}L_{i}, \bar{L}_{k}=\max _{k} L_{k}
$ and $\bar{\boldsymbol{x}}^k \in \mathcal{A}\cap \mathcal{C}^{N}(v^K)$, with $\mathcal{A}$ as in (\ref{eq: consensus}) and $\mathcal{C}$ being the core (\ref{core}). Therefore, it holds that  $\lim_{k \to \infty} \| \boldsymbol{x}^k - \bar{\boldsymbol{x}}^k\| \leq \frac{\delta}{1-\bar{L}_{k}}.$ $\hfill \square$
  \end{theorem}
We provide the proof of Theorem \ref{thm: payoff allocation paracontraction} in Appendix. Note that we are addressing the problem of tracking the core of a dynamic coalitional game, thus the result of Theorem \ref{thm: payoff allocation paracontraction} shows the convergence of the sequence  $(\boldsymbol{x}^k)_{k \in \mathbb{N}}$ to a neighborhood of the core set that depends on the parameter $\delta$ as in Assumption $\ref{asm: Bounded time variations}$, which bounds the variability over time of the coalitional game. For the problem of tracking the core, the variability of the game can be bounded by assuming non-empty intersection of the two consecutive cores, i.e., $\mathcal{C}(v^{k-1}) \cap \mathcal{C}(v^{k}) \neq \varnothing $. Note that, if the game is static, then the iteration in (\ref{main_it_core_general}) converges to the common point in the core set, i.e., the agents employing the algorithm will reach consensus on the core payoff distribution. Thus, the online payoff distribution protocol in (\ref{main_it_core_general}) is asymptotically consistent \cite{Simonetto2020}, which is an important feature of online algorithms.\\ 
\textit{Discussion:}
To use the payoff distribution algorithm in (\ref{main_it_core_general}), each agent requires information on its own bounding set in (\ref{eq: bounding set}) only that can be evaluated using the values of its own coalitions. Thus, this negotiation via bounding sets maintains inter-agent privacy. It is reasonable to assume that the agents have knowledge of their own coalitions. 

We note that the centralized version of online tracking in the context of time-varying convex optimization is presented by Simonetto in \cite{Simonetto2017}. However, centralized methods for tracking a payoff in the core do not capture scenarios of interaction among autonomous self-interested agents. Furthermore, as the core is a set in which different payoffs treat agents differently, a centralized evaluation will demand the trust of agents on the central entity, which is undesirable in many real-world applications, e.g. peer-to-peer energy exchange \cite{raja2021payoff}. Thus, we propose a distributed method in (\ref{main_it_core_general}) that allows agents to autonomously track a core payoff distribution. 

Interestingly, for the class of games (e.g. convex games) where the Shapley value belongs to the core, the online tracking of Shapley value via the iteration in (\ref{main_it_shapley}) implicitly tracks the core and vice versa via the algorithm in (\ref{main_it_core_general}). 
\vspace{-0.3cm}
\section{Real-time market applications}
 In this section, we illustrate numerically the scenarios of two real-time markets, i.e., a forecasting market and a local electricity market, modeled as the dynamic coalitional games. In the first scenario, we present a distributed tracking of the Shapley value for an online data valuation scheme; in the second scenario, we simulate a real-time local electricity market and track the time-varying core payoff as an online market solution. 
 \vspace{-0.5cm}
 \subsection{Collaborative forecasting market}
We simulate the near real-time collaborative forecasting market described in Section \ref{sec: intro} for an application of wind power generation.  {Here, we model a market for trading point forecasts instead of probabilistic forecasts to remain consistent with the most widely adopted practice for wind power prediction \cite{pinson2009probabilistic}.} Normally, wind energy is forecasted for horizons of hours ahead. However, if the wind power penetration in a system  reaches a certain high level, it becomes crucial for the system's security to also
have forecasts with a lead time ranging from 1 to 30 minutes. These short-term to near real-time predictions are required for various operations in the power systems, e.g. by the transmission system operator (TSO) for the continuous balance of the power system, as an input to the (offshore) wind farm controllers, and for the
operation of wind-storage systems providing system regulation \cite{pinson2012very}. Therefore, we design a market-based prediction system for near real-time wind energy forecasting based on online coalitional games. 
 \begin{figure}
\centering
\includegraphics[width=0.85\columnwidth]{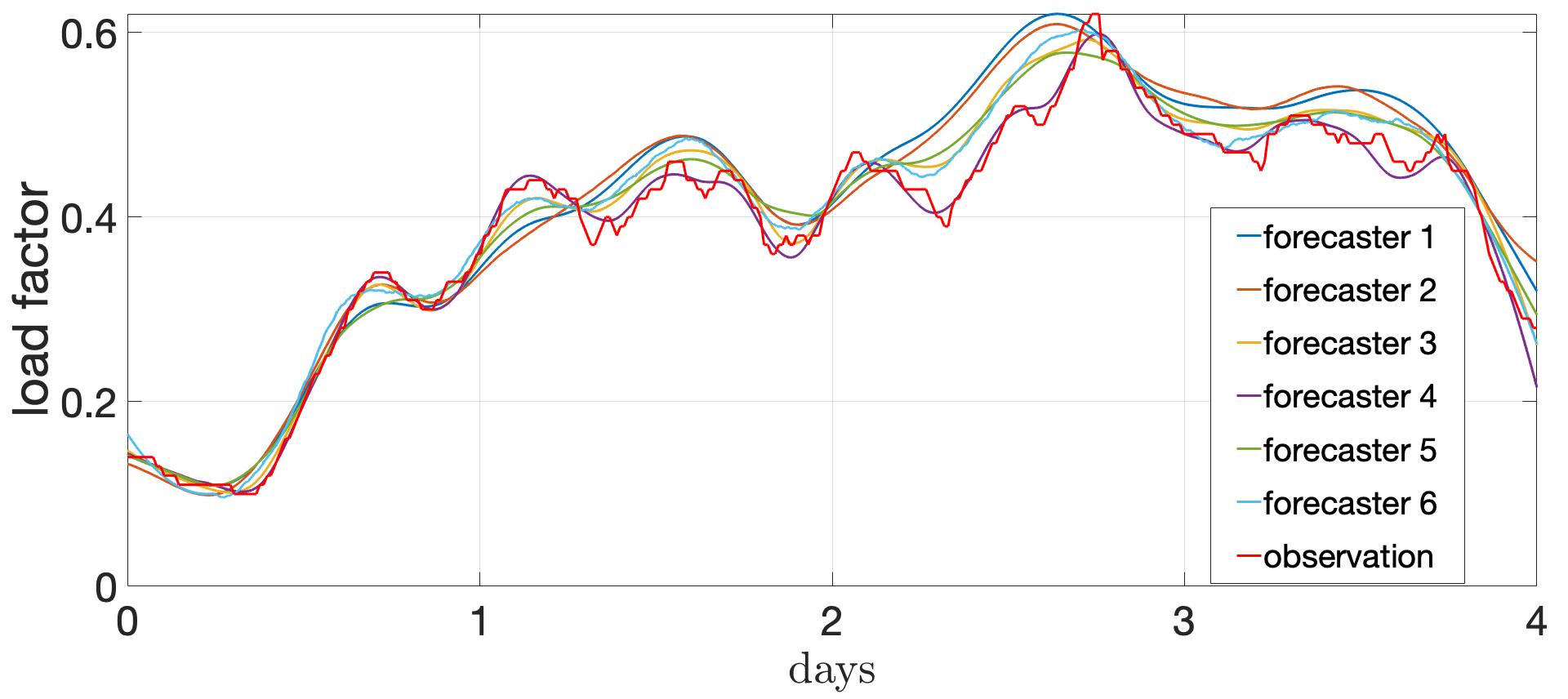}
\caption{Agents' forecasts of wind energy generation with a lead time of 5 minutes and corresponding observation.}
\label{fig: forecasts}
\vspace{-0.5cm}
\end{figure}
\subsubsection{Problem setup}
Consider a client's platform $\mathcal{L}$ (e.g. TSO, wind farm owner, energy trader, etc.) that uses a wind energy forecast to optimise decision-making in highly dynamic environments.  The client organises a collaborative forecasting market with the task of predicting wind energy generation $Y^k$ at time instant $k$ for time $k + m$, where $m$ is on the scale of a few minutes. We consider $N$ forecasters (agents) that register on the client's platform to participate in the near real-time collaborative forecasting market. Each forecaster $i \in \mathcal{I}$ posts a point forecast $f_i^{k+m}$ at time $k$, which is a conditional expectation of $Y^{k+m}$. Then, the client uses linear pooling to evaluate an aggregated forecast $ \hat{f}_\mathcal{I} = \textstyle \sum_{i \in \mathcal{I}}\frac{1}{|\mathcal{I}|}f_i$. After the event occurs and the actual wind energy generation $\omega$ is observed, the client's platform evaluates the quality of the aggregated forecast. Then, the client announces the reward $\phi$ to be distributed among the forecasters according to the quality of their predictions.\\
In the literature, the most widely used criteria to evaluate the quality of forecasts are the so-called scoring rules \cite{gneiting2011making}. For our work, we use absolute error (AE) as a scoring rule which is used for the evaluation of point forecasts. Let the reported prediction by a forecaster $i$, be $f_i$ and  let $\omega$ be  the  actual outcome, then their AE is given as $ \text{AE}_i=|f_i - \omega|.$\\
We can now formulate this collaborative forecasting market as a coalitional game by letting the value of a coalition $S \subset \mathcal{I}$ to be $(1 - \text{AE})$ of its combined forecast, i.e., $v(S) = 1 - |\hat{f}_S - \omega|$. Each forecaster evaluates the values of its own coalitions and utilizes the online protocol in (\ref{main_it_shapley}) to distributedly track the Shapley payoff. The payoff represents the share of each forecaster in the reward evaluated by Shapley value. Note that, we compute the Shapley value as a payoff factor which corresponds to the monetary payoff that an agent will receive.  {The correspondence of payoff factor to monetary payoff is application specific and depends on the gain in the monetary utility of the client because of the collaborative forecast. For instance, in our example of wind energy forecasting, the payoff can correspond to the improvement in utility by optimal operation of combined wind-hydro power plants or by avoiding an imbalance charge in the market. To keep our focus on the market mechanism, we do not consider monetary payoffs and remark that their incorporation would not affect the resulting solution properties. }

\begin{figure}
\centering
\includegraphics[width=0.85\columnwidth]{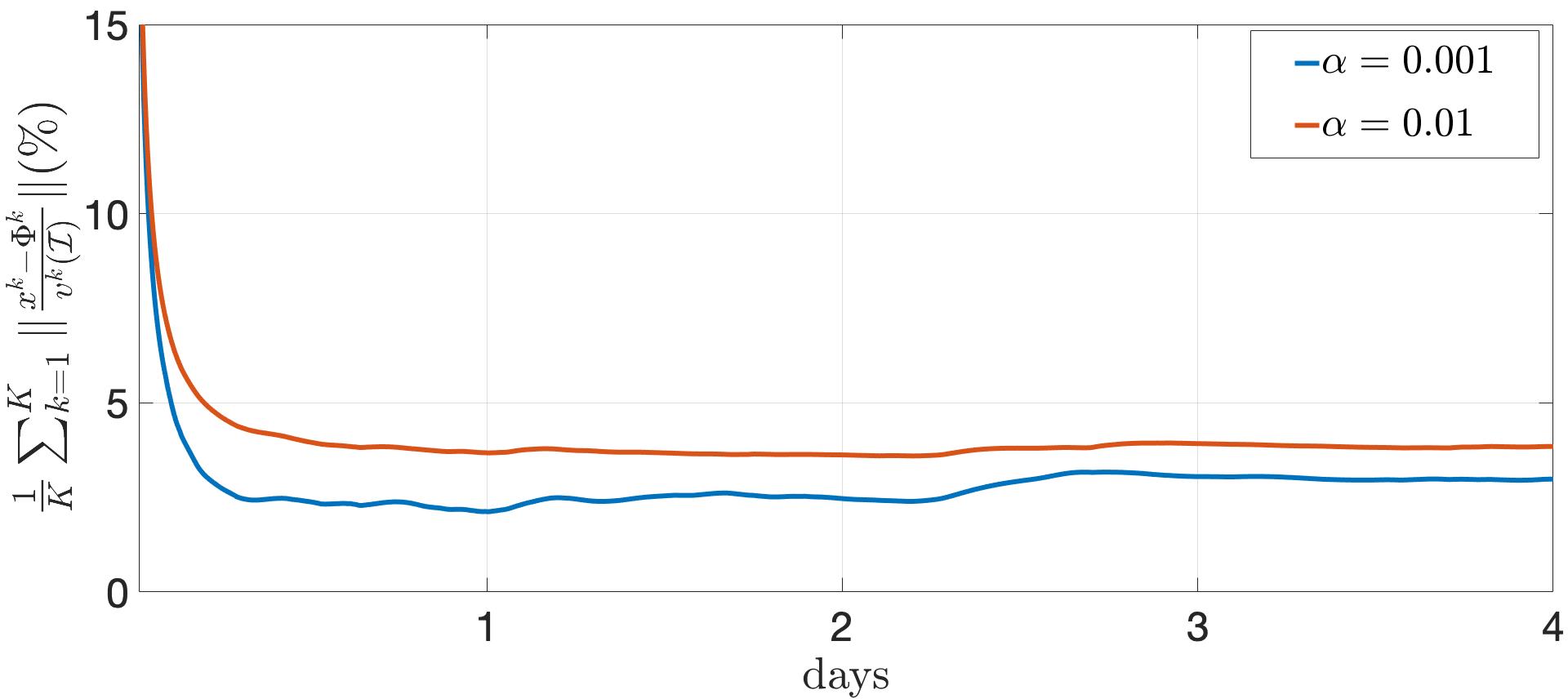}
\caption{Trajectory of mean cumulative tracking error $ \textstyle \frac{1}{K}\sum_{k=1}^K \|\frac{\boldsymbol{x}^k - \boldsymbol{\Phi}^k}{v(\mathcal{I})}\| $, where $\boldsymbol{\Phi}$ is the Shapley allocation.}
\label{fig: cum_avg shapley}
\end{figure}

\begin{figure}
\centering
\includegraphics[width=0.85\columnwidth]{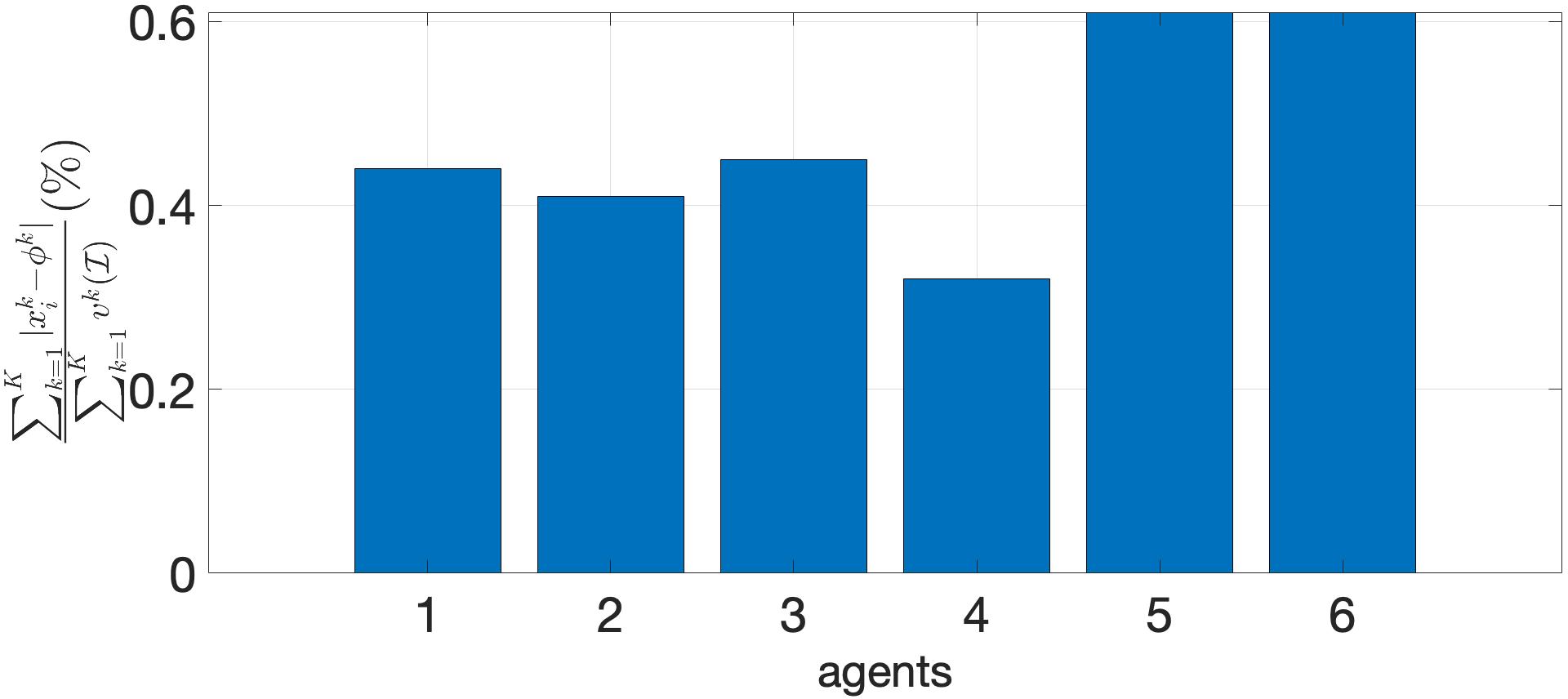}
\caption{Difference in payoff received by online PD and the Shapley payoff $|\boldsymbol{x}_i^k - \boldsymbol{\phi}^k|$ proportional to the total value generated in the market $\sum_{k=1}^K v^k(\mathcal{I})$.}
\label{fig: shapley payoff}
\vspace{-0.5cm}
\end{figure}
\subsubsection{Simulation study}
To illustrate the collaborative forecasting market, we consider that a client sets up a micro market with the task of forecasting wind energy generation in Germany with a lead time of $5$ minutes. Let $6$ forecasters (agents) register at the client's platform for providing the forecast reports.  { Each forecaster posts their prediction of wind power in the form of a point forecast at time $k$ for lead time $k + 5$ minutes. The client then aggregates the reported forecasts to generate a collaborative prediction and utilizes the mechanism in (\ref{main_it_shapley}) for real-time payoff distribution. Let this market run continuously for $4$ days to create a time series.}  Here, we use synthetic data to simulate forecasters' predictions generated using the forecast and actual measurements provided by the Spotrenewables and interpolate it to get the required resolution for the period of 25-28/05/2022. Fig. \ref{fig: forecasts} shows the agents' forecasts and corresponding observations in terms of the capacity factor, i.e., normalized to the theoretical maximum of wind power plant for a $4$-day period. The high accuracy of generated forecasts simulates the near real-time forecasting effect. Next, in Fig. \ref{fig: cum_avg shapley}, we present the tracking performance of our algorithm in (\ref{main_it_shapley}) for different values of $\alpha$.  {We compute the tracking error by evaluating the online payoff and the Shapley payoff (static case) for the market game at each instant $k$ as $ \textstyle \frac{1}{K}\sum_{k=1}^K \|\frac{\boldsymbol{x}^k - \boldsymbol{\Phi}^k}{v(\mathcal{I})}\| $. In words, we report the norm of the normalized difference between the online payoff and the Shapley payoff accumulated over time $K$.} This cumulative tracking error is less than $4 \%$ for both values of $\alpha$. Generally, for short-term to near real-time energy-related markets, the forecast accuracy is high and there is a low variation from one time-step to the next. Thus under such setups, our algorithm shows promising performance. We stress that the tracking performance of our algorithm depends on the dynamics of an underlying problem. Abrupt changes in the value function $v^k(S)$ can increase the tracking error significantly. Fig. \ref{fig: shapley payoff} shows the difference in forecasters' payoff over four days with the Shapley payoff.  

 \subsection{Real-time local electricity market}
 In this subsection, we simulate a real-time local energy trading with an electricity market setup inspired by \cite{guo2021online}. In our proposed setup, the prosumers and consumers participate in a  local electricity market, established within the community, to trade energy internally rather than with a grid. The economic viability of such a market setup is based on the assumption that the buyers value energy higher than the grid's buying price and not more than the grid's selling price. Similarly, the sellers choose their valuation less than the grid's selling price. We note that these assumptions are common in the literature \cite{han2018incentivizing}. Traditionally, electricity markets are organized in a day-ahead setting with some intra-day arrangements for balancing purposes. However, due to uncertainty in RES and consumer load, at the level of a community, the market-clearing so far ahead of delivery can be considerably problematic for the system operator, responsible for system security. One way to mitigate the effect of uncertainty is organizing a market close to the time of delivery.  In this direction,  we design an online market mechanism based on dynamic coalitional game theory for a real-time market model. The  dynamic formulation incorporates an evolving energy demand and RES generation that change with time. In this market setup, ideally, the goal is to maximize the social welfare of the local electricity market and distribute the resulting amount among participants such that the payoff should belong to the core in (\ref{core}). However, in a real-time clearing setup, it is not possible to compute a payoff in the core exactly, thus we track it via the online mechanism in (\ref{main_it_core_general}).
 
\begin{figure}
\centering
\includegraphics[width=0.85\columnwidth]{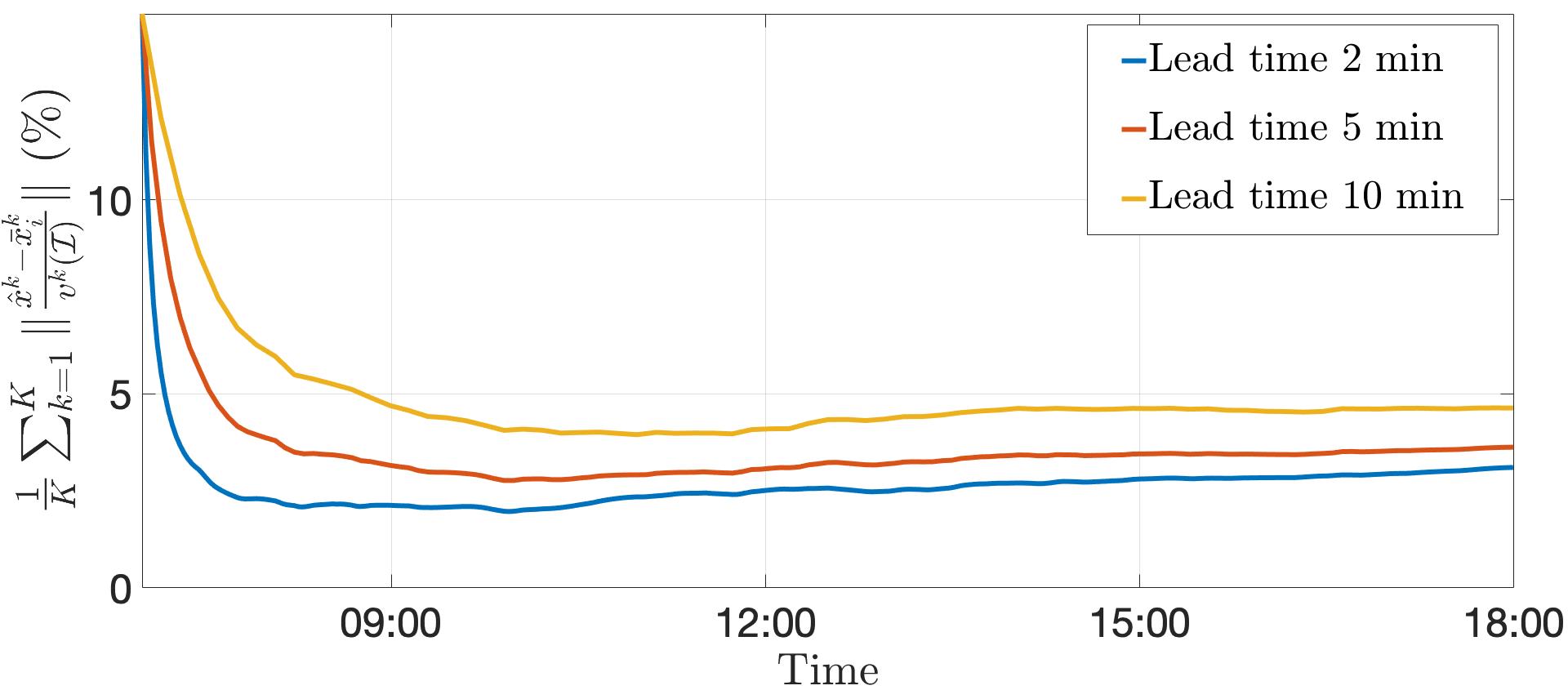}
\caption{Trajectory of mean cumulative tracking error $\frac{1}{K}\sum_{k=1}^K \|\frac{\boldsymbol{\hat{x}}^k - \boldsymbol{\bar{x}_i}^k}{v(\mathcal{I})}\| $, where $\boldsymbol{\hat{x}} = \textstyle \frac{1}{N}\sum_{i \in \mathcal{I}} \boldsymbol{x}_i $ and $\boldsymbol{\bar{x}}_i$ is the core allocation.}
\label{fig: cum_avg}
\vspace{-0.5cm}
\end{figure}
\subsubsection{Problem setup}
We consider a simplified setup with $N$ agents in an energy community $\mathcal{I}$, some equipped with RES generation (prosumers). We compute the coalitional value of each coalition $S \subseteq \mathcal{I}$ for a time instant $k$ by solving a linear optimization problem. At each $k$, an agent either belongs to a set of buyers $\mathcal{S}_b$  or sellers $\mathcal{S}_s$ where, $\mathcal{S}_b \cup \mathcal{S}_s = S$. Let us denote the energy demand or generation of an agent $i$ at a time instant $k$ by $E^k_i$ and the corresponding utility function coefficient by $p^k_i$. Here, we take the utility function coefficient of a seller as negative, i.e., $p_i < 0$ if $i \in \mathcal{I}_s$. We compute the coalitional value $v^k (S)$  for each coalition $S \subseteq \mathcal{I}$ as follows:
\begin{equation}
v^k(S) =
\left\{ \quad
\begin{aligned}
 \max _{\substack{(E_i)_{i \in S}}} & \sum_{i \in S} p_i^k E_{i}^{k}  \\ 
  \mathrm{s.t.} \quad &\: 0 \leq (E_{i}^k)_{i \in S} \leq (\bar{E_i}^k)_{i \in S}  \\
& \sum_{i \in S_s}E_i^k - \sum_{i \in S_b}E_i^k = 0. \
\label{problem setup_PV} 
\end{aligned} 
\right.
\end{equation}
The constraints in (11) show instantaneous generation and consumption limits of sellers and buyers, respectively, and a power balance. We note that only mixed coalitions, i.e., with buyers and sellers, will produce a value, a fact that reduces the computational burden.  At every market instant, each agent computes its bounding set in (\ref{eq: bounding set}) and then proposes a payoff via the online protocol in (\ref{main_it_core_general}).  {To compute their bounding sets, agents need to compute the values of only their own coalitions, which requires information on individual coalitional values. We assume our local market to be established in an advanced paradigm, in combination with futuristic data markets for energy systems like \cite{goncalves2020towards}}. This allows the agents to acquire the information required for computing the bounding sets. At the first instant of the market, agents allocate the whole value  $v(\mathcal{I})$ to themselves, which is in accordance with their rational and self-interested nature. The goal here is to maximize the social welfare of the local electricity market and then distribute the resulting amount among participants such that the payoffs track the core. Note that the unserved demand and unutilized generation will be traded with the grid.

\begin{figure}
\centering
\includegraphics[width=0.85\columnwidth]{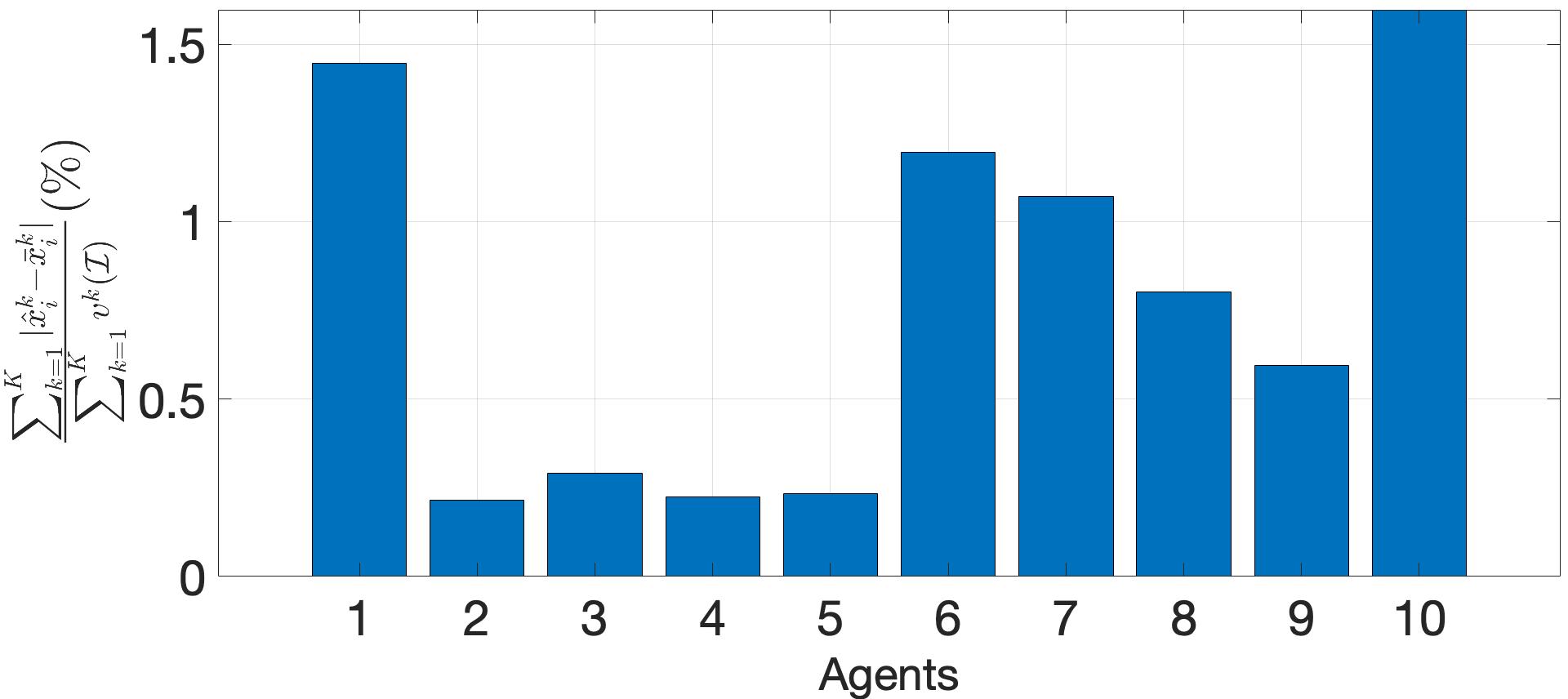}
\caption{Difference in payoff received by online PD and the core payoff $|\boldsymbol{\hat{x}}_i^k - \boldsymbol{\bar{x}_i}^k|$ proportional to the total value generated in the market $\sum_{k=1}^K v^k(\mathcal{I})$, with lead time of five minutes.}
\label{fig: bar payoff}
\end{figure}

\begin{figure}
\centering
\includegraphics[width=0.85\columnwidth]{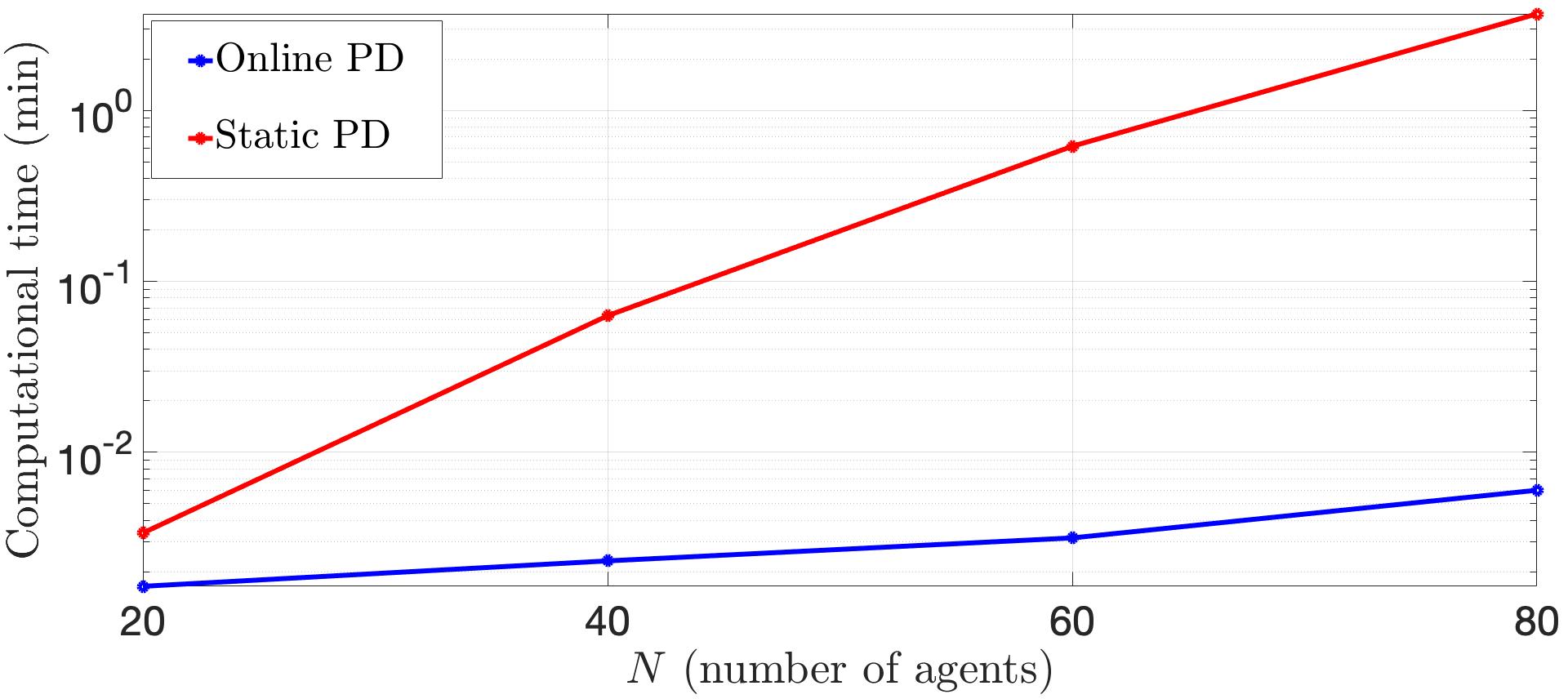}
\caption{Computational time of an agent $i$ for online payoff distribution (PD) and static PD case.}
\label{fig: comp time}
\vspace{-0.5cm}
\end{figure}

\subsubsection{Simulation study}
For the numerical simulation, we consider a small local electricity market of 10 participants where the seller agents are equipped with PV systems and buyers are consumers. We use real data of PV generation and consumer load, recorded at $10$-minute intervals, provided by a smart-grid demonstration project in the UK named Customer-Led Network Revolution (CLNR) \cite{CLNR}. We analyse market-clearing with the lead time of $k + 2, k + 5$, and $k + 10$ minutes. For the $2$ and $5$ minute lead time, we interpolate CLNR's data to achieve the required resolution. Furthermore, we only consider the time slots that have considerable PV generation during the day to demonstrate the effectiveness of our algorithm.  {At each market instance, the seller agents post the energy available to trade and its asking price. While buyer agents post their energy demand and willingness to pay for it. After receiving offers and demands, the participants negotiate to divide the optimal welfare of the market $v^k(N)$ by (\ref{problem setup_PV}).} We note that in an online setting, we track a consensus among agents on a core payoff instead of exact convergence to it.  {Therefore, because of the distributed formulation of our algorithm, the payoff proposals of agents at each market instant can differ and a criterion is required to allocate a mutually agreed payoff.} In this simulation study, we select an average of all proposals to allocate a payoff $\boldsymbol{\hat{x}}$, where $\boldsymbol{\hat{x}} = \textstyle \frac{1}{N}\sum_{i \in \mathcal{I}} \boldsymbol{x}_i$. For evaluating the tracking error, we let the algorithm converge  to a consensus on a core payoff allocation, i.e., $\boldsymbol{\hat{x}}_i = \boldsymbol{\hat{x}}_j$ for all $i,j \in \mathcal{I}$. Finally, in Fig. \ref{fig: cum_avg}, we report mean cumulative tracking error $\frac{1}{K}\sum_{k=1}^K \|\frac{\boldsymbol{\hat{x}}^k - \boldsymbol{\bar{x}_i}^k}{v(\mathcal{I})}\| $ that shows the core tracking capability of the algorithm.  {Since with a lead time of $10$ minutes the market conditions (generation and load) change more from one market-clearing instance to another than with a lead time of $2$ minutes, the tracking error is higher in the former case.} This observation is consistent with the result in Theorem \ref{thm: payoff allocation paracontraction}. Interestingly, for our market setup, the cumulative tracking error is below $5 \%$ even in the 10-minute case. Next, in Fig. \ref{fig: bar payoff}, we present the difference in the payoff of each agent from the core payoff for a lead time of 5 minutes. The difference is at most $1.6 \%$ only, thus supporting the financial viability of an online payoff distribution in real-time markets. Finally, to report a comparison of computational times of online payoff distribution with the static case across the market size, we simulate a time-varying version of the bilateral P2P market presented in \cite{raja2021fair}. In Fig. \ref{fig: comp time}, we show that exactly computing the core payoff is not feasible for fast-paced markets.
\vspace{-0.25cm}
\section{Conclusion}
 {In this paper, we propose a real-time payoff distribution in online coalitional games where the goal is to track a consensus on the payoff distribution solutions, namely, Shapley value and the core. We have shown that an online paradigm of coalitional games provides promising tools for modeling collaborative systems working in environments with fast dynamics, e.g., such as real-time markets. The proposed distributed algorithms based on contraction operators adequately track the payoff distribution solutions. Our examples of local electricity market and collaborative forecasting market show the extent of energy-related applications that can be formulated with our proposed online framework. Numerical simulations illustrate the benefits of our online protocol and show that under the bounded variation in coalitional values a reasonable aggregate difference in online payoff and corresponding exact solutions can be achieved. Thus, online algorithms address the problem of scalability in real-time markets well modeled as coalitional games.\\
Next, we envision a competition platform to test the performance of the proposed online market mechanism and the behavior of participants in practical scenarios. Such a setup should provide useful insights for real-world implementation of the mechanism. An interesting extension of our work would be to incorporate long-term forecasts in the online formulation to provide better performance for events with high volatility.  
}


\appendix
To prove the convergence of iteration in (\ref{main_it_shapley}) and (\ref{main_it_core}), as stated in Theorem \ref{thm: online Shapley allocation} and Theorem \ref{thm: payoff allocation paracontraction}, respectively, we first provide useful results regarding contraction operators.
\begin{lem}[\cite{Simonetto2017}, Thm. 3.1]\label{lemma: contraction convergence}
Let $ \{M^k\}_{k \in \mathbb{N}}$ be a sequence of contraction operators with $ \{L^k\}_{k \in \mathbb{N}}$ as corresponding contraction factors such that $ \mathrm{fix}(M^K)_{k \in \mathbb{N}} \neq \varnothing $. Let Assumption \ref{asm: Bounded time variations} hold. Then, the error norm $|| x^{k} - \bar{x}^{k} ||$ generated by $x^{k+1} := M^k( x^{k})$  converges as: 
$$
\|x^{k}-\bar{x}^{k}\| \leqslant \hat{L}_{k}\|x_{0}-\bar{x}_{0} \|+\frac{1-(\bar{L}_{k})^{k-1}}{1-\bar{L}_{k}} \delta,
$$
where $ \bar{x}^{k} \in \mathrm{fix}(M^k), 
\hat{L}_{k}=\prod_{i=1}^{k-1}L_{i} \text{ and } \bar{L}_{k}=\max _{k} L_{k}.$
$\hfill \square$
\end{lem}
\begin{lem}[Doubly stochastic matrix (\cite{Fullmer2018}, Prop. 5 )]\label{lem: Doubly stochastic matrix}
If $W$ is a doubly stochastic matrix then, the linear operator defined by the matrix $W \otimes I_{n}$ under Assumption \ref{asm: graph} is a paracontraction with respect to the mixed vector norm $\|\cdot\|_{2,2}$. 
\end{lem} $\hfill \square$

\begin{lem}[Composition of a contraction and paracontraction operator  (\cite{bauschke2017convex}, Prop. 4.49)]\label{prop:Composition of operators }
Suppose $T_1:\mathbb{R}^n \rightarrow \mathbb{R}^n$ is a contraction operator and  $T_2:\mathbb{R}^n \rightarrow \mathbb{R}^n$ is a paracontraction with respect to same norm $\|\cdot\|$ and $\mathrm{fix}(T_1) \cap \mathrm{fix}(T_2) \neq \varnothing$. Then, the composition $T_1 \circ T_2$ is a contraction and $\mathrm{fix}(T_1 \circ T_2) = \mathrm{fix}(T_1) \cap \mathrm{fix}(T_2)$. $\hfill \square$
\end{lem}
With these results, we are now ready to present the proofs of Theorems \ref{thm: online Shapley allocation} and \ref{thm: payoff allocation paracontraction}.

\begin{proof}(Theorem \ref{thm: online Shapley allocation})
Let us formulate the iteration in (\ref{main_it_shapley}) as  $\boldsymbol{x}^{k+1} = \boldsymbol{M}^k (\boldsymbol{x}^{k})$ where  $\boldsymbol{M}^k := (\mathrm{Id}-\alpha \nabla f^k)\circ \boldsymbol{W}^k $. Then, by Lemmas \ref{lem: Doubly stochastic matrix} and \ref{prop:Composition of operators } $(\boldsymbol{x}^k)_{k \in \mathbb{N}}$ generates a sequence of contraction operators. For a time-invariant case, i.e.,  $\boldsymbol{z}^{k+1} = \boldsymbol{M} \boldsymbol{z}^k $ by \cite[Lemma 1]{yuan2016convergence} $\boldsymbol{z}^k \to \bar{\boldsymbol{z}}$ as $\|\boldsymbol{z}^k - \bar{\boldsymbol{z}}\| = O(\alpha)$ where $\bar{\boldsymbol{z}}$ is optimizer of the problem in (\ref{eq: shapley optimization}), i.e., $\bar{\boldsymbol{z}} = \boldsymbol{\phi}(v) = \frac{1}{N} \sum_{i \in \mathcal{I}}\boldsymbol{\hat{m}}_i$. Now, in time-varying case, under Assumption \ref{asm: Bounded time variations} the time variation of $\boldsymbol{M}^k$ is bounded, thus the application of Lemma \ref{lemma: contraction convergence} completes the proof.
\end{proof}

\begin{proof}(Theorem \ref{thm: payoff allocation paracontraction})
For the iteration in (\ref{main_it_core_general}), it follows from Lemma \ref{lem: Doubly stochastic matrix} that $\mathrm{fix}(\boldsymbol{T}^k \circ \boldsymbol{W}^k) = \mathrm{fix}(\boldsymbol{T}^k) \cap \mathrm{fix}(\boldsymbol{W}^k) = C^N(v^k) \cap \mathcal{A}$. 
By Lemmas \ref{lem: Doubly stochastic matrix} and \ref{prop:Composition of operators }, the iteration in (\ref{main_it_core_general}) generates a sequence of time-varying contraction operators. Under Assumption \ref{asm: Bounded time variations} the time variation of $\boldsymbol{T}^k$ is bounded, thus the application of Lemma \ref{lemma: contraction convergence} completes the proof.
\end{proof}

\ifCLASSOPTIONcaptionsoff
  \newpage
\fi





\bibliographystyle{IEEEtran}
\bibliography{IEEEabrv,Bibliography}

\vfill


\end{document}